\documentclass[10pt]{iopart}

\usepackage[utf8]{inputenc}
\usepackage[LY1]{fontenc}
\usepackage{txfonts}
\usepackage{graphics}

\begin{document}

  \title[Runout transition and clustering in binary avalanches]
        {Runout transition and clustering instability observed in binary-mixture avalanche
         deposits}

  \author{Roberto Bartali$^1$, Gustavo M Rodríguez Liñán$^2$, Luis Armando Torres-Cisneros$^3$,
          Gabriel Pérez-Ángel$^3$ and Yuri Nahmad Molinari$^2$ }
  \address{$^1$ Facultad de Ciencias, Universidad Autónoma de San Luis Potosí, Salvador Nava 5,
           78290, San Luis Potosí, Mexico.\\
           $^2$ Instituto de Física, Universidad Autónoma de San Luis Potosí, Salvador Nava 5, 
           78290, San Luis Potosí, Mexico.\\
           $^3$ Departamento de Física Aplicada CINVESTAV-IPN, Unidad Mérida, Apartado Postal 73
           Cordemex  97310. Mérida, Mexico.}
  \ead{\mailto{rbartali@uaslp.mx} \mailto{robertobartalim@gmail.com}}

  \begin{abstract}
  Binary mixtures of dry grains avalanching down a slope are experimentally studied in order to
  determine the interaction among coarse and fine grains and their effect on the deposit
  morphology. The distance travelled by the massive front of the avalanche over the horizontal plane
  of deposition area is measured as a function of mass content of fine particles in the mixture,
  grain-size ratio, and flume tilt. A sudden transition of the runout is detected at a critical
  content of fine particles, with a dependence on the grain-size ratio and flume tilt. This
  transition is explained as two simultaneous avalanches in different flowing regimes (a
  viscous-like one and an inertial one) competing against each other and provoking a full
  segregation and a split-off of the deposit into two well-defined, separated deposits. The
  formation of the distal deposit, in turn, depends on a critical amount of coarse particles. This
  allows the condensation of the pure coarse deposit around a small, initial seed cluster, which
  grows rapidly by braking and capturing subsequent colliding coarse particles. For different
  grain-size ratios and keeping a constant total mass, the change in the amount of fines needed for
  the transition to occur is found to be always less than 7\%. For avalanches with a total mass of
  4~kg we find that, most of the time, the runout of a binary avalanche is larger than the runout
  of monodisperse avalanches of corresponding constituent particles, due to lubrication on the
  coarse-dominated side or to drag by inertial particles on the fine-dominated side.
  \end{abstract}
  
  \vspace{2pc}
  \noindent{\it Keywords}: binary avalanches, granular flows, granular clustering, segregation
                           process
  \submitto{\JGE}
  \ioptwocol
  
  \section{Introduction}
  
  Rock avalanches and geophysical granular flows have been broadly studied in order to understand
  the potential risks they represent and the modifications they make on the landscape. They
  mobilize millions of cubic meters of rocks, with sizes ranging from microns up to tens of meters,
  running at velocities that, in some cases, are faster than 100~km/h (Robinson \etal 2015, Clavero
  \etal 2002, Fauque and Strecker 1988, Siebert 1984, Ui 1983). Since their behaviour depends on a
  myriad of parameters, such as particle-size distribution, hardness, textural characteristics,
  friction and restitution coefficients, kind of triggering event, and the environment in which
  they develop, travel, and deposit, they are very complex phenomena, which are difficult to
  predict and/or to be modelled. Any interstitial fluid, like hot gas (such as in pyroclastic
  density currents) or water (like in mudflows and lahars), also plays a fundamental role in the
  flow behaviour, due to lubricating effects, which, in turn, increase the associated risk. For
  this reason, experiments with model particles and simple geometries are performed at laboratory
  scale in order to understand the main features of the avalanching process. In this simplifying
  spirit, monodisperse avalanches and the role that basal friction has on the runout have been
  systematically studied, \emph{e.g.,} by Goujon \etal (2003). They found a minimum basal friction,
  which depends on particle size, and developed a simple model to determine how the geometric
  parameters of the rough base modify the runout of the avalanche.
  
  Several groups have studied the origin of the scaling laws that describe the length and height of
  the deposit as a function of the initial conditions (Pouliquen 1999, Andreotti \etal 2002) and
  coefficients of restitution of materials (Campbell 2006). Other observations relate the distance
  travelled by avalanches with the topography of the terrain or with the presence of model forests
  at the slope change (Yang \etal 2011). Several other mechanisms may affect the mobility of
  avalanches, for example, dynamic fluidization (Huerta et al 2005), fluid-pore-pressure increment,
  or ball-bearing effects induced by the presence of fine particles within large inter-particle
  voids (Pacheco-Vázquez and Ruiz-Suárez 2009, Hungr and Evans 2004). Several hypotheses regarding
  the mechanisms that lead to long runouts and the mobility of rock avalanches are summarized in
  Charrière \etal (2015) and references therein. Increased mobility of avalanches containing large
  and fine grains is also observed in numerical simulations (Linares-Guerrero \etal 2007). In
  summary, particle-particle and particle-base interactions have been the object of thorough
  research to shed some light on the problem of how the multiple factors of a real avalanche are
  related with the final developed deposit and its runout, height and length, in order to better
  predict the risks they could represent.
  
  During their development, avalanches present an irregular flow characterized by an intermittent
  motion (stick-slip) and formation of density waves. Van Gassen and Cruden (1990) developed a
  model in which they describe the increments in runout due to intermittent deposition of a dry
  granular flow. Later on, some stick-slip motion of the flow down a slope was reported by Bartali
  \etal (2015) They developed a medium-size, densely instrumented experimental flume (Bartali
  \etal 2012, Bartali \etal 2015) and found evidence of this intermittent behaviour in
  polydisperse, dry granular flows and an increment of the affected area through the ejection of
  ballistic projectiles.
   
  Stick-slip motion will very likely produce reminiscent features in the final deposit. Seeking
  for these reminiscences, Paguican (2014) performed analogous model experiments aimed to
  reproduce and understand the origin of hummock formation and the dynamics of spreading and
  aggregation. The intermittent motion causes concomitant faulting and sliding zones that will
  finally form the characteristic elevations and mounds scattered on top of the deposition area.
  In field observations, intriguing hummocks, constituted mainly by coarse grained rocks and
  situated beyond the avalanche front, have been described and interpreted by Clavero \etal
  (2002).
   
  On increasing complexity, mixtures of two particle sizes in different proportions have also been
  investigated. In those cases, ubiquitous segregation occurs in different forms: levees, vertical
  segregation by kinetic sieving, inertial dispersion of ballistics, etc. Several authors report a
  non-monotonic behaviour of the runout (mobility) of avalanching masses as a function of
  fine-particle content (Phillips 2006, Yang 2015, Moro 2010, Goujon \etal 2007). Likewise,
  Kokelaar \etal (2014) used rough, irregularly shaped materials mixed with smooth, rounded grains
  in order to describe how both materials interact with each other and with the base while they
  flow down the slope. They observed a levee formation that guides the avalanche by channelling or
  confining the flow of fine grains by the so-called ``outline effect''. Similarly, Kokelaar \etal
  (2014) observe the growth of the runout up to a maximum occurring at a critical proportion of
  fine particles.
   
  In summary, a critical fine-particle content is responsible for the maximum mobility of the
  avalanche by lubricating interactions among large particles and between large particles and the
  base, and it has been well established for binary avalanches. All these reported data suggest
  that the runout is a continuous function of the fine content. However, in this work we report
  that each size component has its own very well differentiated runout. During the flow, the
  avalanche spreads out and segregates by size, continuously changing the effective composition of
  the mixture in different locations. The avalanching mass is thus a non-homogeneous mixture of
  two dynamically changing species along the channel until it completely stops. The different
  runouts of each species would lead, in the case of a complete separation during the flow, to a
  two-folded runout value or a discontinuous transition of the runout as a function of
  fine-particle content, never observed or reported before.
   
  The main objective of this work is to systematically address the runout value in binary
  avalanches with enough contrast in grain-size species, finely exploring the behaviour of the
  avalanche runout around the critical value where the runout reaches its maximum, modifying the
  fine-particle content in steps as small as 2.0\%. Changes in runout account for a different
  rheological behaviour and could help in developing risk-management strategies and in providing a
  more accurate interpretation of past geological flows from their deposits. The purpose of our
  experiments is to understand the interactions between dry, irregular, natural particles running
  down a flume. Particle-particle interactions (collisions and friction) may be changing
  dynamically, so the microscopic (grain-level) behaviour, must influence the macroscopic
  (flow-level) behaviour and, hence, the velocity of the flow and its energy dissipation per unit
  time. The final runout and the morphology of the deposit must be a direct consequence of the
  interactions of all the grains.
   
  \section{Experimental setup}
  
\begin{figure}[htb]
    \begin{center}
      \noindent\resizebox{\columnwidth}{!}{\includegraphics{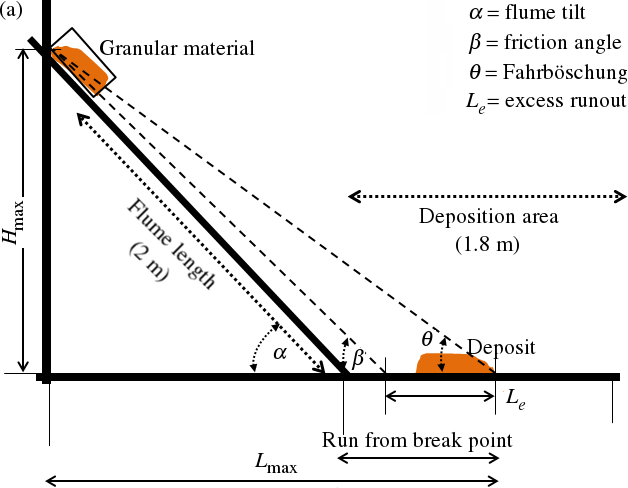}}\\
               \resizebox{\columnwidth}{!}{\includegraphics{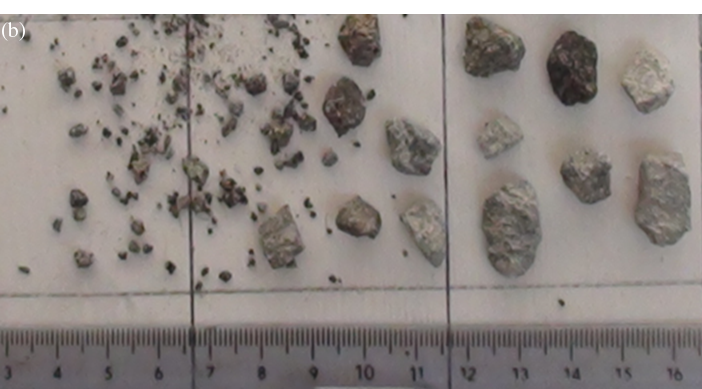}}
      \caption{\label{Fig_Exp}(a) Schematic diagram of the experimental flume. The parameter
               definitions are given in the text when used. The flume tilt, $\alpha$, can be set
               from 0$^\circ$ to 45$^\circ$. (b) Example of the irregular volcanic particles used
               to perform the experiments.}
    \end{center}
  \end{figure}

  We built a laboratory medium-scale (2.5-m long, 15-cm wide), variable-tilt (0$^\circ$ to
  45$^\circ$) flume (figure~\ref{Fig_Exp}a), followed by a 1.8-m-long and 1.2-m-wide, horizontal
  deposition area. Both the flume and the deposition area were built using 15-mm thick, smooth,
  medium-density fibreboard (MDF) sheets. The flume base was covered by a hard, smooth Formica sheet
  in order to reduce the basal friction, while the deposition area was covered with a vinyl paint
  to obtain some roughness. The walls in both sections consisted of a 15-cm-height, 6-mm-thick,
  transparent, tempered glass. An open-frame, metallic structure (50~cm $\times$ 50~cm $\times$
  2.5~m) was employed to hold the flume at the desired tilt. A manual gate was placed 2~m up-stream
  (from the break in slope) in order to accumulate the granular material and start the avalanche
  with zero velocity. This gate was connected to a micro-switch that allowed the automatic trigger
  of a high-resolution ($2208 \times 1244$~px), high-frame-rate (120~fps) video camera (Sony
  HDR-HR150).
  
  We also painted a square grid, 5~cm in side, over the base of the deposition area to allow a
  better scaling and measurements of the runout and the morphology of the avalanche deposit. For
  all the experiments, we used irregular, rough, natural particles collected in the same area to
  ensure uniformity of density (2.6~g/cm$^3$), morphology, and textural features
  (figure~\ref{Fig_Exp}b). This material (andesite clasts from a deposit of a block and ash flow at
  Nevado de Toluca volcano in Mexico) was carefully sieved in order to obtain grain-sizes classes,
  based on the Wentworth (1922) scale, from $-4$~phi (16~mm) to 4~phi (0.0625~mm) mean grain
  diameter. We also sieved the granular material to obtain intermediate grain-size classes:
  $-1.5$~phi (3~mm), $-0.5$~phi (1.5~mm) and 1.5~phi (0.375~mm). The total mass obtained for each
  granular class was 4~kg. We used standard sieves, so a monodisperse set of grains means that it
  contains all the grains that pass through the mesh. In order to ensure a better grain-size
  selection, the loads of grains were sieved twice.
  
  \section{Method}
  
\begin{figure}[htb]
    \begin{center}
      \resizebox{\columnwidth}{!}{\includegraphics{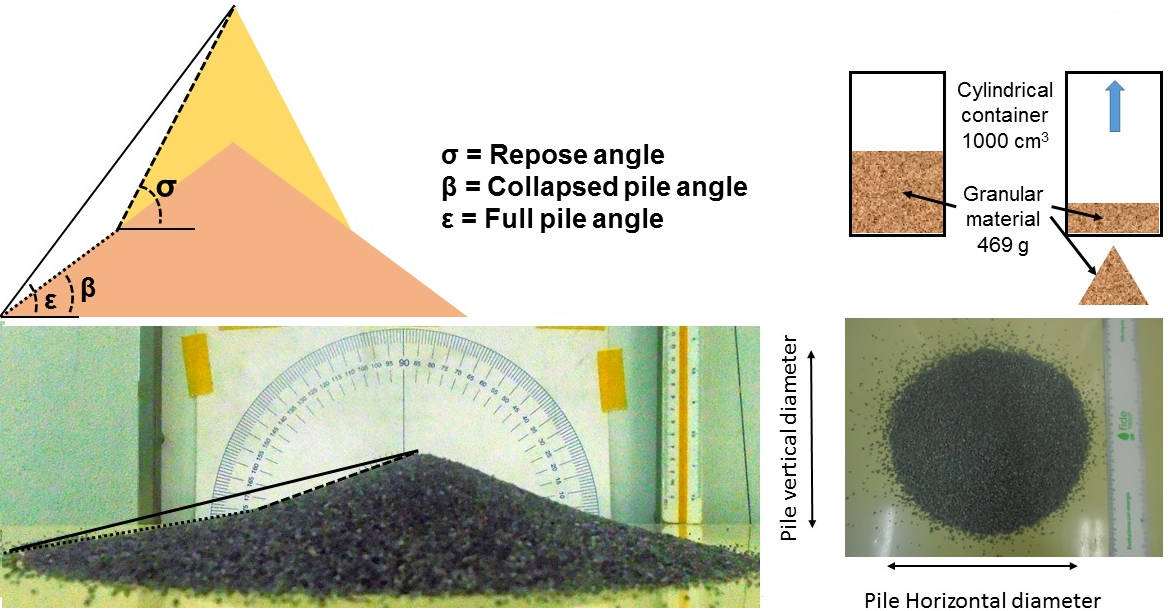}}
      \caption{\label{Fig_Pile} Experimental setup and definition of the angles measured after the
               formation of the pile.}
    \end{center}
  \end{figure}

  Different granulometric classes were characterized by determining their repose angle and static
  friction coefficients. For doing so, collapsing-granular-column experiments and tilted table
  tests were performed ten times for each particle size. For repose angle determination, an
  8-cm-radius tube was filled with 469~g of granular material, and the tube was slowly and
  vertically removed in such a way that the granular column collapses into a conical mound. The
  true shape of the mound is shown in figure~\ref{Fig_Pile}, were three different angles are
  defined. The repose angle is represented by $\sigma$, whereas $\beta$ and $\varepsilon$ represent
  respectively the smooth slope of the material that collapses after the pile formation and the
  full-pile angle related to $\sigma$ and $\beta$. Static friction coefficients were determined by
  gluing a layer of granular material on the base of a wood slab, which was weighted and checked
  for the critical angle at which it starts sliding down over a smooth metallic plate.
  
\begin{figure}[htb]
    \begin{center}
      \resizebox{\columnwidth}{!}{\includegraphics{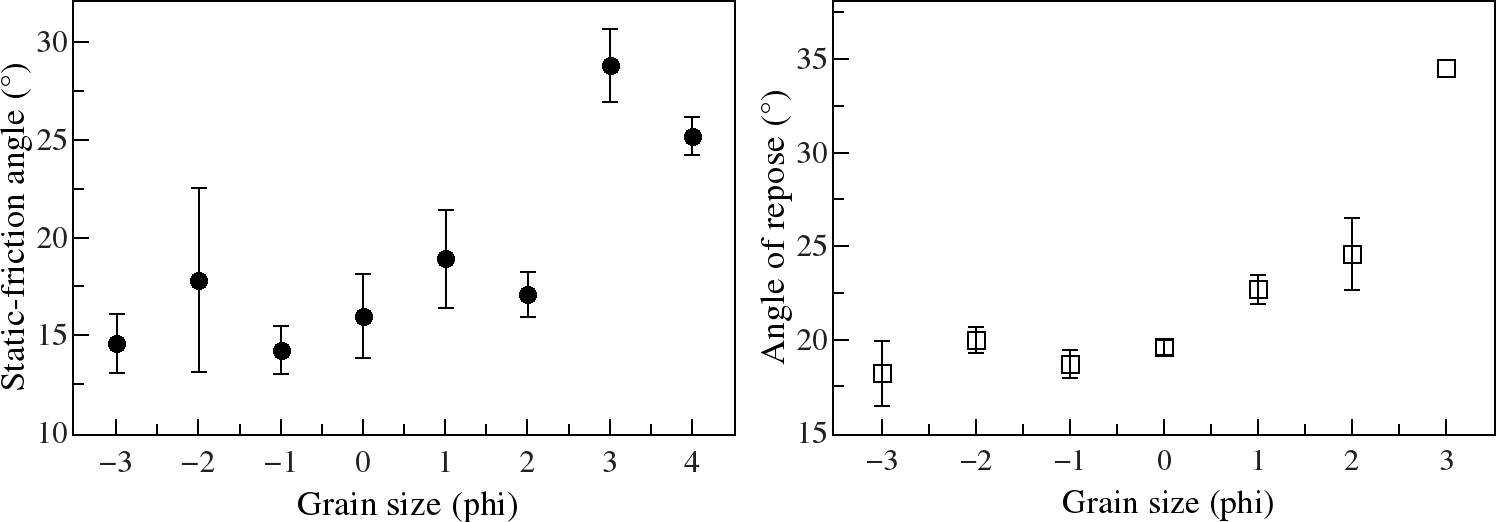}}
      \caption{\label{Fig_Static} 
              (a) Static-friction angle in tilted-table experiments as a function of the 
                  granulometric class phi. (b) Repose angle as a function of the granulometric
                  class.}
    \end{center}
  \end{figure}

  Static-friction angles and repose angles are plotted in figure~\ref{Fig_Static} as a function of
  the granulometric class, phi, showing an increasing trend as the particle size diminishes. 

  We performed experiments aimed to understand the behaviour of dry avalanches containing only
  particles of two different sizes mixed together. We combined different grain classes maintaining
  the total mass of the granular material constant (4~kg) and then changing the relative proportion
  of each class. The first run was always performed with the coarsest grained sample, while the
  second was the finest grained one, both cases having 4~kg of particles, as stated before. This
  methodology rules out the possibility of contamination with fine particles of larger grains
  masses. Before each experiment, we slowly poured all the grains (previously sieved) in a bucket,
  slowly turning with a ladle, trying to get the best mixture. We performed experiments with the
  flume tilted at 32$^\circ$ and at 37$^\circ$, maintaining, in all cases, the deposition area
  horizontally. The set of experiments is summarized in table~\ref{Table_Exp}.
  
  \begin{table}[htb]
  \begin{center}
    \caption{\centering\label{Table_Exp} Set of the different experiments performed.}
    \begin{tabular}{ccc}
      \br
      Flume tilt & Coarse grains & Fine grains\\
      (degrees)  & phi (mm)      & phi (mm)   \\
      \mr
      32         & $-3$ (8)      & 2 (0.25)   \\
      32         & $-2$ (4)      & 2 (0.25)   \\
      32         & $-1$ (2)      & 2 (0.25)   \\
      37         & $-3$ (8)      & 1.5 (0.375)\\
      37         & $-1.5$ (3)    & 1.5 (0.375)\\
      \br
    \end{tabular}
  \end{center}
  \end{table}
  
  The video camera was placed in front of the flume and perpendicular to the deposition area, in
  order to have almost all the flume and part of the deposition area in the field of view. The
  camera was triggered automatically by a micro-switch connected to the gate, which was used to
  start the avalanche. Several markers on the flume and on its walls and the grid painted on the
  deposition surface helped to measure the velocity and the horizontal displacement of the granular
  flow body and that of scattered high-velocity-ejected particles. After each experiment, we sieved
  the granular material, weighted the amount to be used in the next run, and mixed all the grains
  again. In order to ensure that the experimental conditions were almost the same for all the
  experiments, we carefully cleaned the flume and the deposition area, because fine and very fine
  grains, in small quantities, increase the mobility of the avalanche, as we will show in the next
  section. The flume was cleaned each time, first with a brush and, if very small particles were
  found on the base or walls, with a wet cloth or running water. If this was the case, the flume
  was dried with a hairdryer. The granular material used in each experiment was always from the
  same initial set of 4~kg, in order to maintain the same textural and morphological
  characteristics.
  
  All the materials selected for the flume (Formica and tempered glass) and the deposition area are
  almost static-free. We performed several tests in order to find out which grain size was the most
  affected by electrostatic attraction, sprinkling the flume at several tilt angles. It turned out
  that only 62-\textmu{}m (4~phi) and finer particles were significantly adhered to the base and
  the walls of the flume; consequently, we chose to not perform experiments with those grain sizes.
  
  \section{Results and discussion}
  
  In the following sections, we will discuss the results of the experiments and compare them in
  order to show that they are consistent. We will show that segregation of particles depends on
  particle size and content ratio. Finally, we propose that granular clustering is the origin of
  the deposit morphologies observed experimentally. We also propose that granular clustering can be
  an alternative or an associated process involved in the hummock-formation mechanism for which
  there are several explanations in the literature (Peguican \etal 2014, McColl and Davie 2011,
  Clavero \etal 2002, Glicken 1996, Siebert 1984, Ui 1983).
  
  \subsection{General description of the deposit}
  
\begin{figure}[htb]
    \begin{center}
      \resizebox{\columnwidth}{!}{\includegraphics{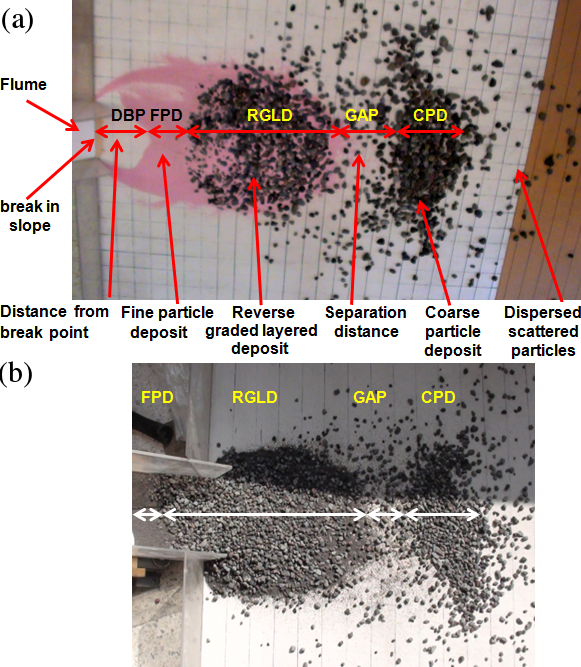}}
      \caption{\label{Fig_Depos} (a) Example of a binary avalanche deposit consisting of 2.2~kg of
               8-mm (black) particles mixed with 1.8~kg of 0.375-mm (pink) particles. The flume
               tilt was set to 37$^\circ$. (b) Example of a binary avalanche deposit consisting of
               2.5~kg of 4-mm (large, gray) particles mixed with 1.5~kg of 0.25-mm (small, gray)
               particles. The flume tilt was set to 32$^\circ$. There are clearly three different
               deposits, due to the segregation processes occurring during the avalanche, the
               collision against the break in slope, and the travel of particles on the horizontal
               deposition area.}
    \end{center}
  \end{figure}

  In figure~\ref{Fig_Depos}a, a typical deposit of a binary avalanche is shown. In this case, the
  coarsest particles measure 8~mm (55\% by mass), while fine grains are 0.375~mm (45\% by mass).
  Clearly, the deposit is segregated, presenting a region of only fine particles (pink), a region
  with a mixture of coarse and fine particles (the finest at the bottom and the coarsest on top)
  and a region in which just coarse particles are deposited (dark gray). In some cases, a gap opens
  up between the mixed and the coarse-particle regions. If the mobility of the avalanche is not
  large enough to overpass the break in slope and to deposit on the horizontal deposition area, the
  fine-particle deposit and part of the mixed-particle deposit remain on the flume, as can be seen
  in figure~\ref{Fig_Depos}b. In this deposit, the largest particles have a mean diameter of 4~mm
  (62.5\% by mass), and the finest ones measure 0.25~mm (37.5\% by mass).

  The characteristic lengths of the different parts of the deposit (see figure~\ref{Fig_Depos}) are
  defined as the ``distance from break point'' (DBP), which is measured from the break point to the
  tail of the fine-particle deposit. The fine-particle deposit length (FPD) refers to the distance
  from DBP to the point in which a mixture of particles can be seen. At this point starts the mixed
  deposit, which actually consists of two perfectly reverse-graded layers (RGLD), and its length is
  measured to the point in which particles are no longer in mutual contact. Between the
  mixed-particle deposit and the coarse-particle deposit (CPD) there are scattered particles
  forming a gap (GAP), consisting of a low-density cloud of coarse particles. This low-density
  cloud of scattered particles ends where the density of particles suddenly increases, forming a
  cluster of coarse-particle deposit (CPD). Outside this last region, particles are scattered and
  form a large V-shaped cloud. The distance from the break in slope to the tip of the
  coarse-particle deposit will be called hereafter ``run from the break point'' (RBP). All these
  lengths are measured on the horizontal deposition area along the symmetry axis of the flow and
  will be very important at describing the complex phenomenology of the granular, binary-mixture
  avalanches.
  
  \subsection{Detailed deposition sequence}
  
\begin{figure}[htb]
    \begin{center}
      \resizebox{\columnwidth}{!}{\includegraphics{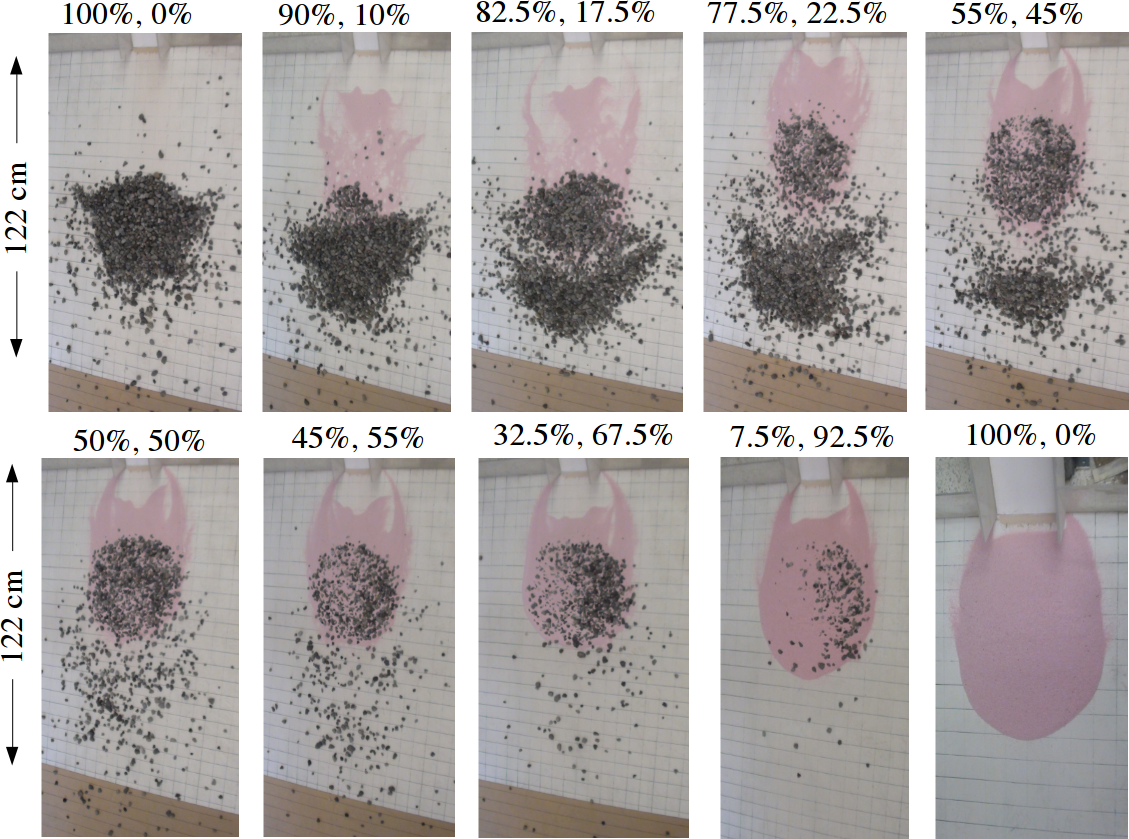}}
      \caption{\label{Fig_Seq-3_1.5} Sequence of deposits created by the binary avalanche, containing
               8-mm (gray) and 0.375-mm (pink) particles. Fine-particle content in the mixture
               increases from left to right. A dramatic change in the deposit morphology is clear
               when the fine-particle content increases from 45\% to 50\%. The deposit suddenly
               retreats. The grid on the horizontal deposition area is 5~cm $\times$ 5~cm.}
    \end{center}
  \end{figure}

  In figure~\ref{Fig_Seq-3_1.5}, a sequence of pictures of the final deposits for ten different fine
  contents is shown. The avalanche contained a mixture of 8-mm particles (large, black) and
  0.375-mm fine particles (pink). These experiments were performed with a flume tilt of 37$^\circ$.
  The fifth pictured deposit, corresponding to 55\% of large particles and 45\% of fine clasts, is
  the one shown previously in figure~\ref{Fig_Depos}a. 

  It is interesting to observe that, when the mixture contains 55\% of large particles and 45\% of
  fine particles (see the fifth and sixth pictures in figure~\ref{Fig_Seq-3_1.5}), the deposit shows a
  bi-modal structure; in some cases, there is a large cluster of coarse particles at the distal
  part of the deposit and, in other cases, a very small one. This double behaviour can be obtained
  by changing the content of fines by a factor of 3\% or less.
  
\begin{figure}[htb]
    \begin{center}
      \resizebox{\columnwidth}{!}{\includegraphics{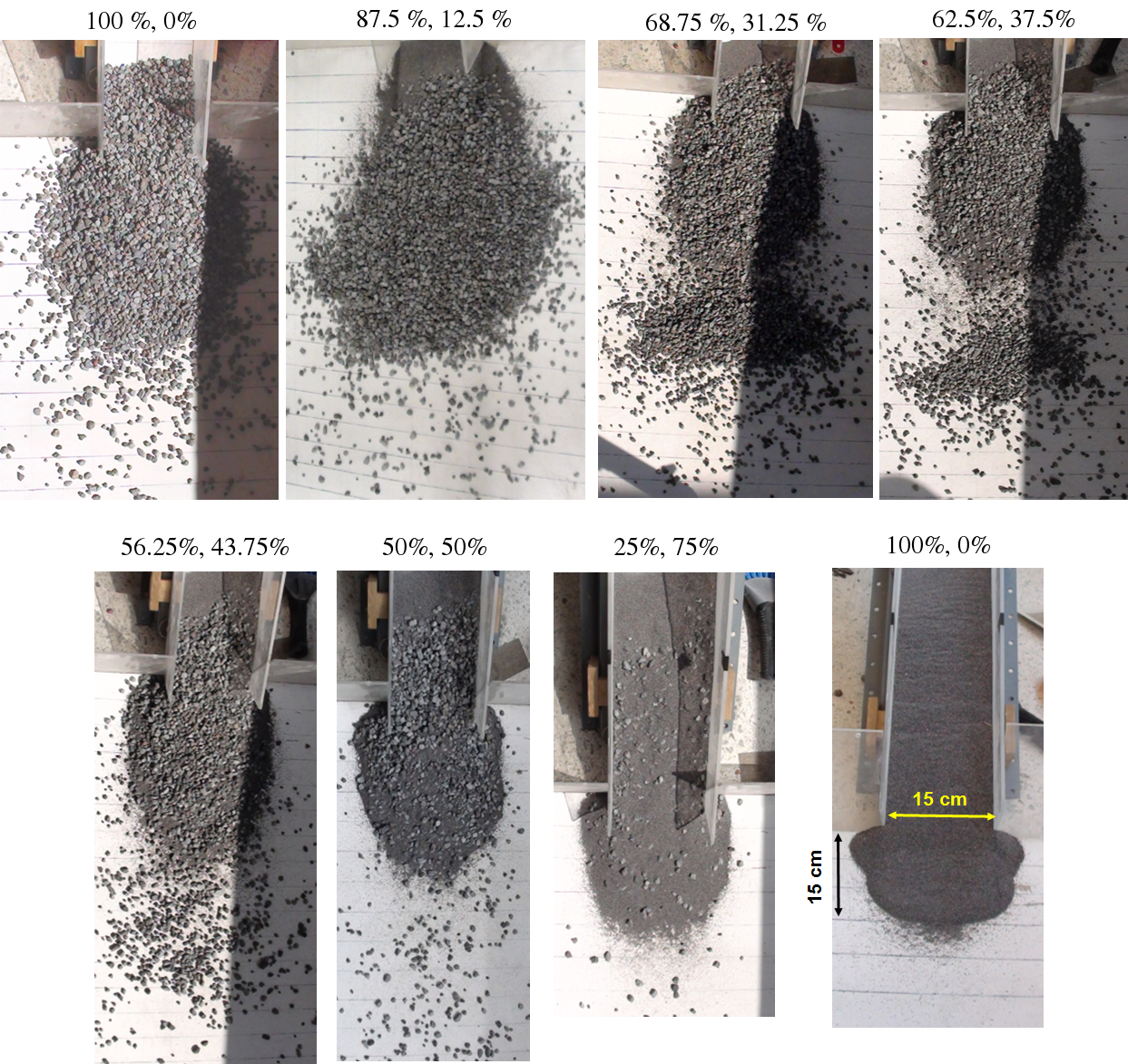}}
      \caption{\label{Fig_Seq-2_2} Sequence of deposits created by the binary avalanche containing
               4-mm (large, gray) and 0.25-mm (small, gray) particles. The fine-particle content in
               the mixture increases from left to right. A dramatic change in the deposit
               morphology is clear when the fine-particle content is increased from 43.75\% to
               50\%. The deposit suddenly retreats. The grid on the horizontal deposition area is
               5~cm $\times$ 5~cm.}
    \end{center}
  \end{figure}

  The sequence of deposits shown in figure~\ref{Fig_Seq-2_2} was produced by an avalanche of 4-mm
  particles (dark gray) mixed with 0.25-mm ones (light gray). The content of fine particles in
  figure~\ref{Fig_Seq-3_1.5} and figure~\ref{Fig_Seq-2_2} increases from left to right, growing
  from 0\% to 100\% by mass. The sudden change in the deposit morphology and, hence, the retreat of
  the runout, is obtained when the fine-particle content in the mixture changes from 43.75\% to
  50\%, when the large cluster at the distal part of the deposit suddenly retreats (see the fifth
  and sixth pictures in figure~\ref{Fig_Seq-2_2}).

\begin{figure}[htb]
    \begin{center}
      \resizebox{\columnwidth}{!}{\includegraphics{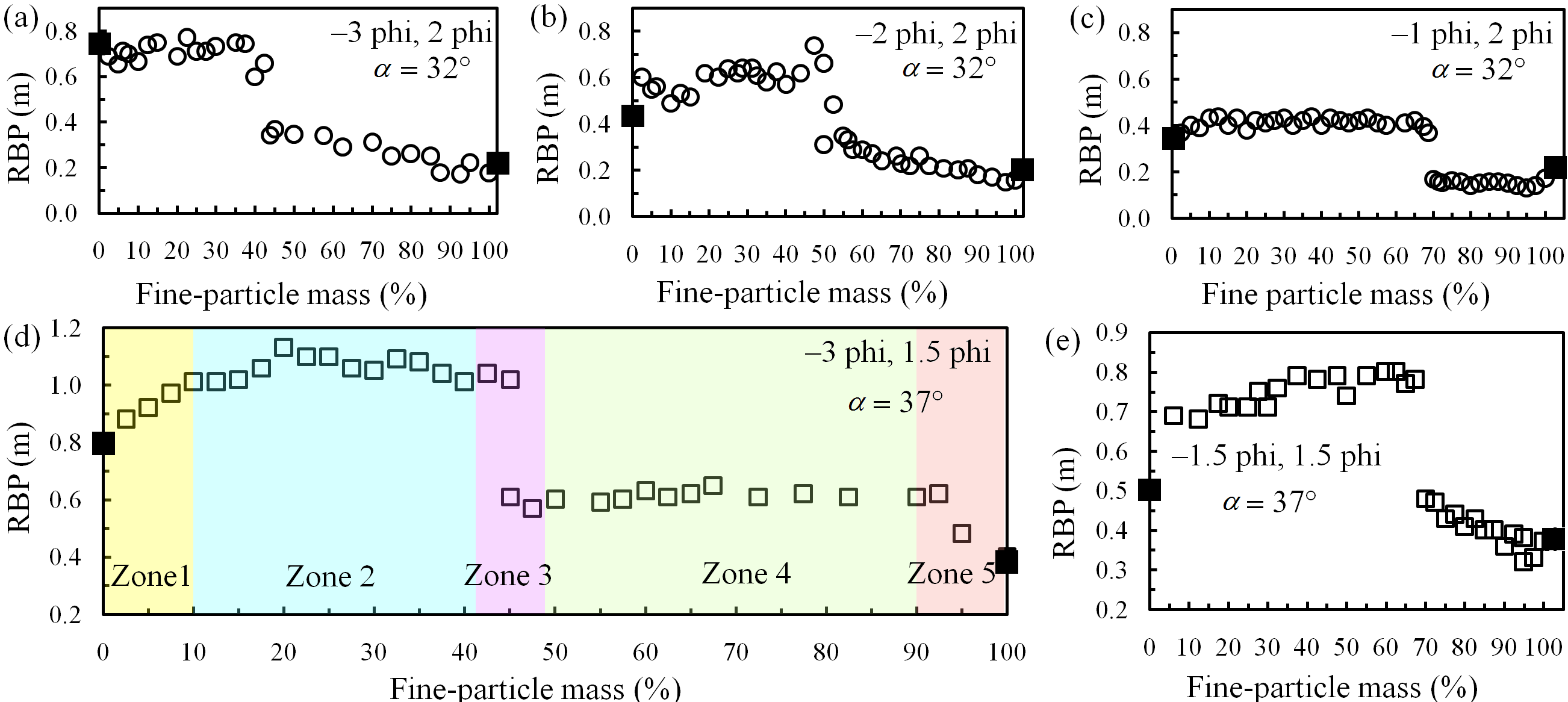}}
      \caption{\label{Fig_RBP} Total distance travelled by the avalanche on the horizontal
      deposition area—``run from break point'' (RBP)—plotted as a function of the fine-grain
      content (\%). Three results—(a), (b), and (c)—with the flume tilted 32$^\circ$ and two
      other—(d) and (e)—for the flume tilted 37$^\circ$ are shown. In all cases, a sudden change in
      the RBP length is observed. The amount of fines needed to produce this transition depends on
      the grain-size ratio in the granular flow mixture. Five different zones can be recognized in
      all plots. For clarity, they are highlighted in plot D: Zone 1, RBP increment; Zone 2, RBP
      almost constant; Zone 3, RBP transition; Zone 4: RBP almost constant; Zone 5, RBP decrement
      or almost constant. The black filled square (on each plot) is the run from break point of the
      large-particle monodisperse avalanche (left side) and that of the small particles (on the
      right side).}
    \end{center}
  \end{figure}

  From the analysis of figures~\ref{Fig_Seq-3_1.5} and \ref{Fig_Seq-2_2} and all the experiments
  listed in table~\ref{Table_Exp}, we get the different plots shown in figure~\ref{Fig_RBP}, in
  which the run from break point (RBP) is depicted as a function of fine-particle content. It can
  be seen that, for a null fine content, the coarse particles form a single, large cluster
  surrounded by scattered particles towards the front. When a few percent of fines are added to the
  mixture, the deposit separates into three clearly defined zones: an only-fine-particle deposit, a
  mixed deposit and an only-coarse-particle deposit. It is worth to note that the mixed deposit
  always consists of two reverse-graded layers, where fine grains are at the bottom and coarse
  particles are on top of them.
  
  An increment in the run from break point (RBP) is observed, and its value depends on the
  grain-size ratio and the flume tilt. This trend continues until the RBP reaches an almost
  constant value. As more fine particles are added to the mixture, the pure-fine deposit is more
  clearly defined and the coarse-particle deposit separates from the mixed deposit, leaving a gap
  full of scattered particles in between. Further addition of fine particles induces the dilution
  of the distal, large cluster containing only coarse particles into one or more small clouds of
  particles (small clusters).
  
  A small (around 5--7\%) increment in the fine-particle content in the mixture produces a sudden
  reduction in the RBP, and both the gap and the clusters of coarse particles disappear. If the
  fine-particle fraction is still increased, the few coarse particles in the flow are not able to
  overcome the fine deposit, thus getting stuck over what would be, otherwise, the fine deposit,
  forming a mixed deposit. However, some coarse particles abandon this last deposit to conform the
  scattered cloud of particles far from the massive deposit. The trend observed in the three cases
  for which the flume was tilted 32$^\circ$ is similar to that when the flume tilt was set to
  37$^\circ$. For this reason, we decided to show only one result for each flume tilt and discuss
  them in general terms.
  
  It should be pointed out that avalanches constituted only of fine particles are deposited mainly
  on the slope, while those constituted of coarse particles are deposited on the horizontal
  deposition area, consistently with the viscous-like and inertial regimes described above. In
  fact, the abrupt change in slope could induce the development of sudden internal passive
  pressures, due to air compression during the deposition process. 

  It is clear that a small percentage of fine particles increases the runout and, hence, the
  horizontal travel (RBP) of the avalanche before it settles down and deposits. We attribute this
  behaviour to lubrication produced by fine particles among large particles and between the
  avalanche body and the flume surface. On the other hand, a large amount of fine particles acts as
  a ``sand-trap'' for coarse ones. Regardless of the grain size, they always get stuck inside the
  massive avalanche and the deposit of fine particles. A large particle colliding against thousands
  of smaller particles loses suddenly almost all its kinetic energy. As the radii fraction tends to
  unity, the contrast in the flow regimes is reduced, the separation of the deposits become more
  difficult, and the whole behaviour will be the one of a monodisperse avalanche.
  
\begin{figure}[htb]
    \begin{center}
      \resizebox{\columnwidth}{!}{\includegraphics{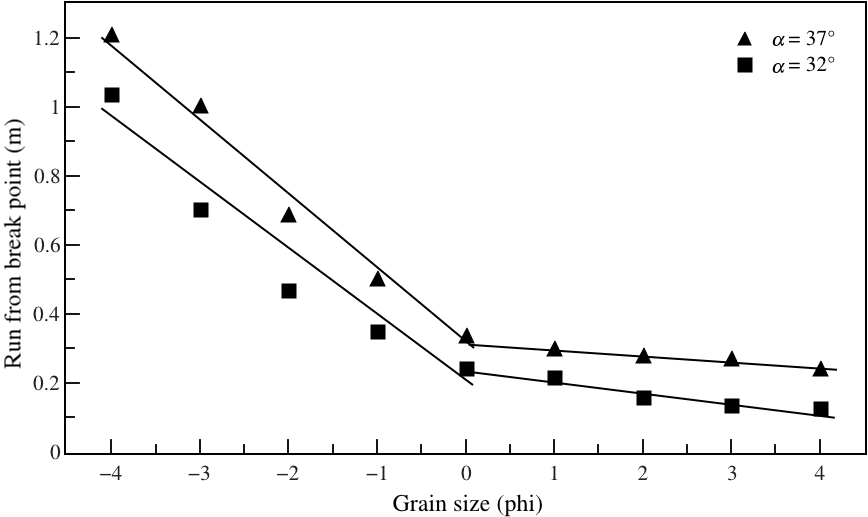}}
      \caption{\label{Fig_RBP-phi} Run from break point (RBP) plotted as a function of grain size
               for monodisperse avalanches. Each point is the mean value after running five
               experiments. The mass of grains in each avalanche is 4~kg. A kink can be seen in
               both trends at a grain size of 1~mm. The RBP for particles larger than 1~mm
               increases proportionally to grain size, while for particles smaller than 1~mm it
               remains almost constant. This plot shows a dramatic change in the behaviour of the
               avalanche from an inertial regime (particles larger than~1 mm) to a viscous-like
               regime (particles smaller than 1~mm), as discussed in text.}
    \end{center}
  \end{figure}

  We should highlight that we have selected mixtures with coarse particles larger than 1~mm and
  fine particles with less than 1~mm, mean diameter. This is because the behaviour of monodisperse
  avalanches containing particles larger than 1~mm is very different to that corresponding to
  granular flows containing particles smaller than this size. We have experimentally determined
  (see figure~\ref{Fig_RBP-phi}) that the coarse-grained ($> 1$~mm), monodisperse-avalanche travels
  in an inertial-like regime, and its RBP (and the corresponding runout) grows proportionally to
  grain size, showing an increment five times larger, or more, for a 16-fold increment in particle
  diameter. 

  For particles larger than 1~mm, the kinetic energy acquired at the expense of the
  potential-energy loss is higher than the energy lost by friction and collisions. On the other
  hand, monodisperse avalanches containing grains smaller than 1~mm behave similar to a
  viscous-like flow, maintaining a much slower increase (they shows a less-than-twofold increment
  for a 16-fold increment in particle diameter) in the runout and RBP values, regardless of the
  grain size. In this case, the energy lost by dissipative processes (friction and collisions) is
  higher than the kinetic-energy increment due to their movement down-slope. This is because, as
  grain size is reduced for the same mass, the number of particles increases as the cube of their
  radius, so the number of mutual collisions and contacts between particles also increases at the
  same proportion.
  
\begin{figure}[htb]
    \begin{center}
      \resizebox{\columnwidth}{!}{\includegraphics{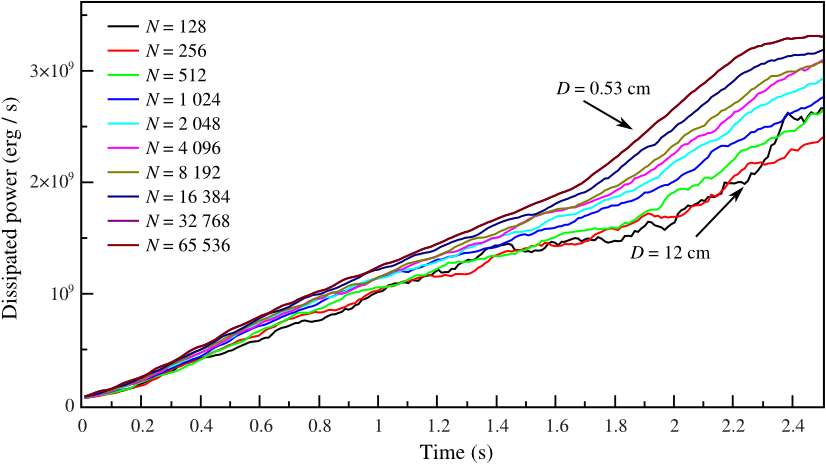}}
      \caption{\label{Fig_Sim} Instantaneous dissipated power of the whole simulational 2D
               avalanche of dimers as a function of time for different number of dimers, keeping
               the total avalanching mass constant.}
    \end{center}
  \end{figure}

  We have further tested the previous hypothesis by performing 2D molecular-dynamic simulations of
  dimers, consisting of overlapped discs, and accounting for the instantaneous dissipated power as
  shown in figure~\ref{Fig_Sim} (simulation details will be published elsewhere). In this figure,
  the total, instantaneous dissipated power is plotted as a function of time for different number
  of particles, $N$, maintaining the total mass constant. Time axis spans from the triggering of
  the avalanche to the moment at which the avalanche front reaches the deposition zone. As it can
  be seen, the dissipated power is always larger for smaller grain sizes, in agreement with the
  aforementioned hypothesis. Having a larger number of particles implies more collisional events
  and, hence, more power dissipated by finer grains. This is consistent with the experimental
  results shown in figure~\ref{Fig_RBP-phi}, in which runout decreases monotonically with
  decreasing grain size.
  
  In figure~\ref{Fig_Static}, the repose angle as a function of grain size—a static measure—shows
  consistency with the dynamic response shown by RBP in the sense that friction is more important
  for small grains than it is for large ones. However, the dynamic response measured trough RBP
  involves the acceleration stage, as well as the deceleration produced at the break in slope,
  which is a much more complex phenomenon.
  
  \subsection{Comparison of experimental results}
  
  In this section we show the results obtained after running the avalanches with a change of the
  fine-particle content of 2.5\% in each step (in almost all cases). This finely tuned exploration
  of the run from the break point as a function of the fraction of fine particles in the mixture
  was made because a sudden transition was detected during the experiments. As can be seen in
  figures~\ref{Fig_Seq-3_1.5}, \ref{Fig_Seq-2_2}, and \ref{Fig_RBP}, as the proportion of fine
  grains increases from zero to one, several behaviours can be noted. In all plots in
  figure~\ref{Fig_RBP}, the run from the break point (RBP) of mixed avalanches is plotted. Five
  different regions can be observed in all plots depicted in figure~\ref{Fig_RBP} but, for clarity,
  we have highlighted them only in figure~\ref{Fig_RBP}d. In some cases, differences among regions
  are too subtle to be discerned. However, the main transition reported in this paper is always
  present and is notoriously sharp. 
  
  First, at very small fine percentages, around 5–15\%, after the first coarse-grained,
  monodisperse-avalanche RBP value (the first black square), a growing tendency of the RBP as a
  function of fine-particle content is recorded (depicted as zone 1), followed by a much slower
  growth rate (zone 2), until a sudden step-function-like fall of the RBP occurs (zone 3). After
  this sudden transition, a slow decreasing of the RBP follows (zone 4) and, finally, for all
  cases, except the one shown in figure~\ref{Fig_RBP}d. This tendency continues until the value of
  the RBP of the fine-grained, monodisperse avalanche is reached, which is always less than or very
  similar to the last RBP of a mixture. This tendency of the RBP occurs when the proportion of fine
  particles is larger than 90\% by mass (zone 5). The observed transition in zone 3 produces larger
  changes in the RBP for larger particle-size ratios and occurs at smaller fine-particle contents
  for larger size ratios, as it can be seen in table~\ref{Table_RBP}. The fourth column in
  table~\ref{Table_RBP} shows that the critical content of fine grains shifts inversely with the
  grain-size ratio for a constant flume tilt. It is worth to note that the transition width is, in
  all cases, close to a 7\%, or less. This variation in fine-particle content of the mixture was
  observed for experiments performed at all flume tilts.
  
  \begin{table*}[htb]
    \begin{center}
      \caption{\centering\label{Table_RBP} Run from break-point transition for all experiments.}
      \lineup
      \begin{tabular}{ccccc}
        \br
        Flume tilt & Transition  particle diameter  & $\Delta$~RBP & Mean fine mass transition content & Fine content transition width\\
        (degrees)  & (ratio)                        & (m)          & (\%)                              & (\%) \\
        \mr
        32 & 32  & 0.41 & 43 & 7\\
        32 & 16  & 0.43 & 48 & 7\\
        32 & 8   & 0.25 & 68 & 5\\
        37 & 21.3& 0.42 & 45 & 3\\
        37 & 8   & 0.31 & 68 & 7\\
        \br
      \end{tabular}
    \end{center}
  \end{table*}
  
  The separation between mixed- and coarse-grain deposits increases linearly with the fine-particle
  content until it falls abruptly again to zero (the two deposits can no longer be distinguished).
  The run from the break point is increased due to the separation of the avalanche into two fronts:
  an inertial one (coarse) and a more viscous-like one (mixed). The sudden transition occurs at the
  point in which the content of fine particles is so large (there is a small number of coarse
  particles) that coarse particles get trapped between the fine ones, and the few ones that can
  overcome the fine-particle deposit cannot form a well-defined deposit. The beginning or the end
  of a deposit is always measured from the point in which the density of scattered particles grows,
  forming large ($> 1/3$ of the flume width) clusters, allowing the stability of superimposed
  particles in two or more layers. The way in which the tip and tail of the deposits are defined is
  critical to understand the discontinuous character of the observed transition.
  
  The addition of few coarse grains to a mixture composed mostly of fine particles only traps these
  coarse grains within the massive fine-particle deposit. The avalanche behaviour is now dominated
  by the viscous-like nature of the flow of fine particles. Further addition of coarse particles
  provokes that some of these much more inertial grains can overcome the deposit of fine and mixed
  particles, which can be found randomly scattered beyond the main deposit, forming a disperse
  cloud. As this process of increasing the coarse fraction is continued, more and more particles
  are added to the coarse-particle cloud, until they start clustering together by inelastic
  interactions.

  The density of the scattered cloud of coarse particles increases until a critical density is
  reached and large clusters form. The first cluster stops the particles coming from the avalanche
  and the just-formed coarse deposit can grow as more coarse particles are added to the mixture.
  This is an inertial deposit, and the subsequent behaviour of the run from the break point will be
  dominated by inertial particles. In this way, one may say that the discontinuous transition of
  the RBP at a critical content of fine grain is actually the separation of the deposit into two
  main deposits: the one of mixed particles and the cluster of scattered, coarse particles that
  condense to form the coarse-grain deposit at the very front. Before the onset of clustering there
  is not a well-defined coarse deposit, and just a diffuse cloud of randomly distributed, coarse
  particles can be identified. Thus, the massive front of the deposited material is composed of
  coarse particles trapped by the fine-grain deposit.
  
  At this point, let us mention, for the sake of clarity, a few words on clustering and
  inelastic-collision-driven transitions in granular matter. ``Inelastic collapse'' (Goldhirsh and
  Zanetti, 1993) is a phenomenon in which simulations of granular gases cooling down by inelastic
  collisions seems to stop. The apparent stall of the simulation occurs because when two particles
  approach, sandwiching a third particle, they cause this third particle to perform a diverging
  number of collisions per unit time. The collision routine is thus invoked a diverging number of
  times but the relative positions of the particles involved in the formation of this cluster evolve
  very slowly as if the simulation were halted. On the other hand, “granular clustering”, closely
  related to inelastic collapse, is the phenomenon in which the restitution coefficient depends on
  the relative collision velocity (as in the case of real grains) and has been reported in
  quasi-two-dimensional granular gases as responsible for segregation in compartmentalized granular
  gases (Olafsen and Urbach 1998, Sapozhnikov \etal 2003, Bordallo-Favela \etal 2009, Perera-Burgos
  \etal 2010, Mikkelsen \etal 2002, Mikkelsen \etal 2005). In the clustering instability, the power
  dissipated in the collisions grows as the inter-particle separation decreases, because the number
  of collisions diverges when they come closer. In this sense, as more particles cluster together,
  it is easier for them to trap particles from the cloud within the cluster. Thus, clustering is a
  self-enhanced process.
  
  The formation of two different deposits separated by a gap, shown in figures~\ref{Fig_Seq-3_1.5}
  and \ref{Fig_Seq-2_2}, could be explained as a consequence of a clustering process. This process
  produces the second well-defined deposit transversal to the flow direction of the avalanche, due
  to the fact that local stresses are of compressive nature. In other words, no extensional
  stresses are possible due to discontinuous granular interactions among grains in this horizontal
  deposition area. This clustering-driven formation of a separated deposit and small, dispersed
  clusters, in our lab-scale experiments, recalls the hummocks observed in natural rock avalanches.
  
  It should be remarked that avalanche-runout increments due to fine particles have been reported
  previously (Kokelaar \etal 2014, Goujon \etal 2007, Phillips \etal 2006). However, in our case,
  lubrication or bearing-ball effects, levee formation, etc., do account for just a small RBP
  increment until the deposit separates into two clearly defined deposits. In the case of a binary
  mixture, there are in fact two avalanches competing against each other.
  
  In our experiments, granulometric classes corresponding to inertial and viscous-like regimes were
  chosen to configure each mixture, in order to discern the effect the viscous-like flow of fine
  particles has on the inertial flow of coarse particles and vice versa. This competition leads to
  the prevalence of one behaviour over the other, depending on the relative content of each
  species, conferring the RBP a dual behaviour. On the one hand, if the inertial avalanche
  dominates, the RBP shows a very large increment, and the deposit can split into two parts: a
  distal, massive cluster of only coarse particles and a proximal, massive deposit consisting of
  only fine particles at the closest distance to the break in slope, followed by a layered,
  reverse-graded, mixed part of the deposit. However, if a gap does not form, the structure of the
  deposit remains the same with three very well differentiated zones: a pure-fine, proximal one; a
  reverse-graded, mixed, intermediate one; and purely coarse-particle, distal zone forming a single
  deposit. On the other hand, if the fine-grain, viscous-like avalanche dominates, the trend of RBP
  is towards diminishing. There is only one massive deposit with just two well defined zones: the
  proximal one of only fine particles attached to the reverse-graded mixed zone; the pure coarse
  grained deposit is absent.
  
  Our findings show that dominance is critical, in the sense that there is no equilibrium state as
  a result of viscous-like-versus-inertial interactions. The transition is always abrupt within a
  7\% (or less) variation of fine content. Additionally, for fixed-total-mass experiments (4~kg),
  we have found that the RBP of a binary avalanche is always greater than (or very similar to) the
  RBP of each corresponding monodisperse avalanches, depicted as a filled black square in
  figure~\ref{Fig_RBP}. On the coarse-dominated side (inertial regime) before the transition, fine
  particles within interstices, formed by large particles, lubricate the interaction among large
  particles or between large particles and flume surfaces. On the other side, where fine particles
  are dominant (the whole behaviour is viscous-like), the RBP is larger than the one of only fine
  particles, due to the drag exerted by large inertial particles on the fine, ``viscous-like''
  particles.
  
  The formation of the distal deposit depends on a critical quantity of coarse particles, which
  allows the condensation of the deposit around a small initial cluster (a ``seed''). This seed
  grows rapidly by braking and capturing subsequent coarse colliding particles. Note that now we
  are discussing in terms of coarse-particle content, instead of fine-particle content, because the
  distal cluster will form or not depending only on the density of scattered particles in the
  disperse cloud at the very front of the deposit.

\begin{figure}[htb]
    \begin{center}
      \resizebox{\columnwidth}{!}{\includegraphics{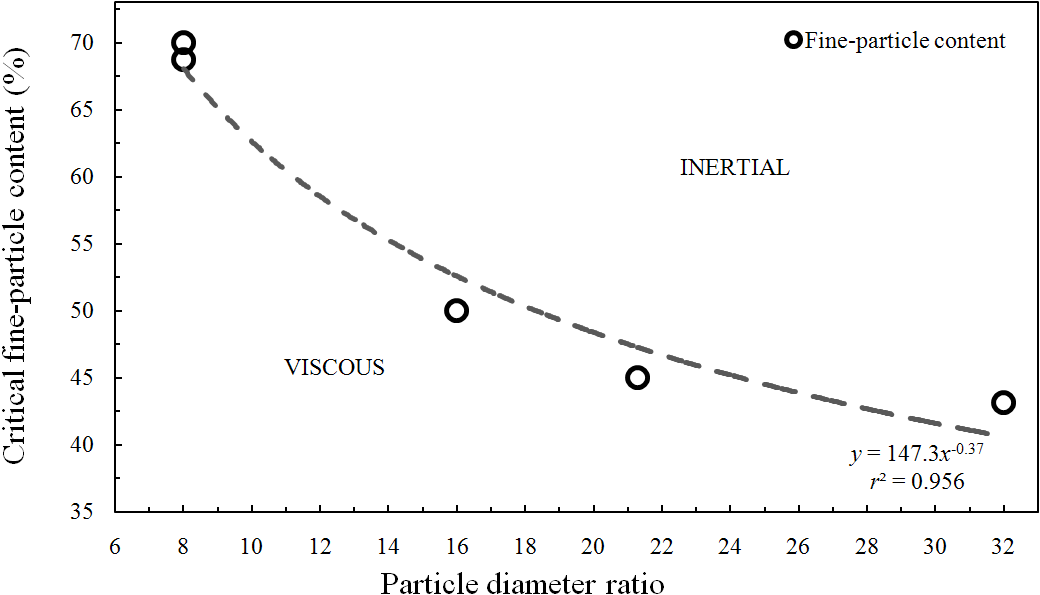}}
      \caption{\label{Fig_PowLaw} Critical amount of fine particles (\%) needed to obtain the
               transition observed as a function of particle-diameter ratio. This line is also the
               border line for the viscous-like-to-inertial regime of the avalanche.}
    \end{center}
  \end{figure}

  By plotting the particle-diameter ratio as a function of critical fine-particle content, a
  power-law behaviour is obtained, as shown in figure~\ref{Fig_PowLaw}. The critical content of
  fine particles diminishes as the particle-diameter ratio grows. This is due to the fact that a
  small quantity of large particles is not enough to constitute a well-defined, coarse-grain
  deposit, because the majority of them are scattered ahead of the massive deposit. This implies
  that the formation of the coarse-grain cluster is not possible until the coarse-particle density
  in the disperse cloud reaches a critical value that allows the formation of large clusters.
  
\begin{figure}[htb]
    \begin{center}
      \resizebox{\columnwidth}{!}{\includegraphics{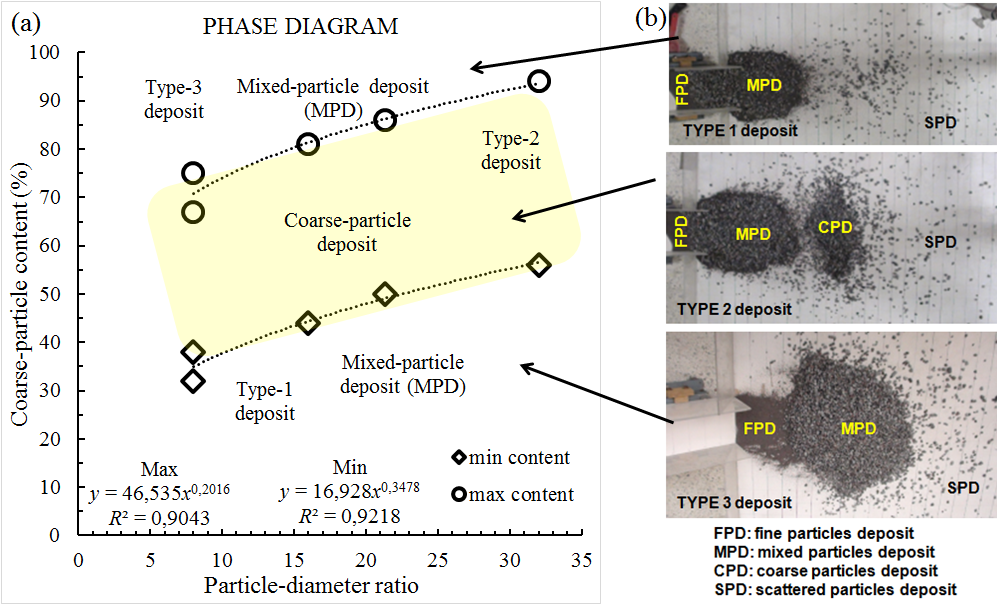}}
      \caption{\label{Fig_Phase} (a) Phase diagram showing the coarse-particle amount needed to
               obtain a coarse-particle deposit (CPD) as a function of particle-diameter ratio. A
               CPD is only allowed (for each particle-diameter ratio) inside the highlighted
               region. Outside this region, the deposit only shows a fine-grain deposit (FPD)
               closer to the break in slope and a mixed-grain deposit (RGLD) in front of it. (b)
               Example of deposits where it can be seen that the coarse-particle deposit forms only
               for a certain amount of coarse particles in the mixture. This quantity is
               highlighted and is a function of the particle-diameter ratio.}
    \end{center}
  \end{figure}

  In figure~\ref{Fig_Phase}, we show a phase diagram, corresponding to the five experiments
  reported in this paper. It can be observed that a deposit conformed by the main four sections of
  the deposit (FPD, RLGD, GAP, and CPD) can only be obtained by a well-defined amount of coarse and
  fine particles in the original mixtures, and it is a function of the particle-diameter ratio. For
  a coarse-particle amount below and above those critical values, the final deposit is conformed
  only by the FPD and the RGLD. In all cases, a small amount of large particles is scattered beyond
  the tip of the massive, distal end of the deposit.
  
  Remarkably, for mixtures in which two deposits form, (a coarse-grained, distal one, and an
  inversely graded proximal one) the morphology of the whole deposit strikingly resembles the
  hummocky deposits of the Parinacota debris avalanche in northern Chile, first described by
  Clavero \etal (2002). As described by these authors, distal hummocks are placed ahead the
  avalanche front and are constituted mainly by coarse-grain rocks that show angled (instead of
  rounded) shapes and collision marks that we attribute to the clustering process we just
  described.
  
  \subsection{Mobility}
  
  Mobility of rock avalanches is often measured using the ratio $H/L$. The height of fall, $H$, is
  measured vertically from the crest of the initial avalanche mass to the lowest point of reach,
  whereas the length of fall, $L$, is measured horizontally from the crest of the avalanche to the
  most distal point of reach (denoted as $H_{\max}$ and $L_{\max}$ in figure~\ref{Fig_Exp}). The
  Fahrböschung angle, $\theta$\/ (Shreve 1968, Hsu 1978), is defined as $\theta = \tan^{-1}
  (H_{\max}/L_{\max})$. We decided to characterize the mobility of our avalanches in this way
  because the travel angle, $\alpha_G$, defined as the angle between the centre of gravity of the
  initial mass and that of the deposit, is much less used, since it is very difficult to be
  determined in natural deposits (Bowman \etal 2012).
  
  In our case, since the deposit sometimes splits into two well-differentiated sequential deposits,
  the Fahrböschung angle is always measured from the rear of the initial granular pile to the tip
  of the most distal condensed or clustered front, which is the tip of the coarse particle deposit,
  CPD—when the deposit is split—or the MDP in all other cases.
  
  \begin{figure*}[p]
    \begin{center}
      \resizebox{0.723\textwidth}{!}{\includegraphics{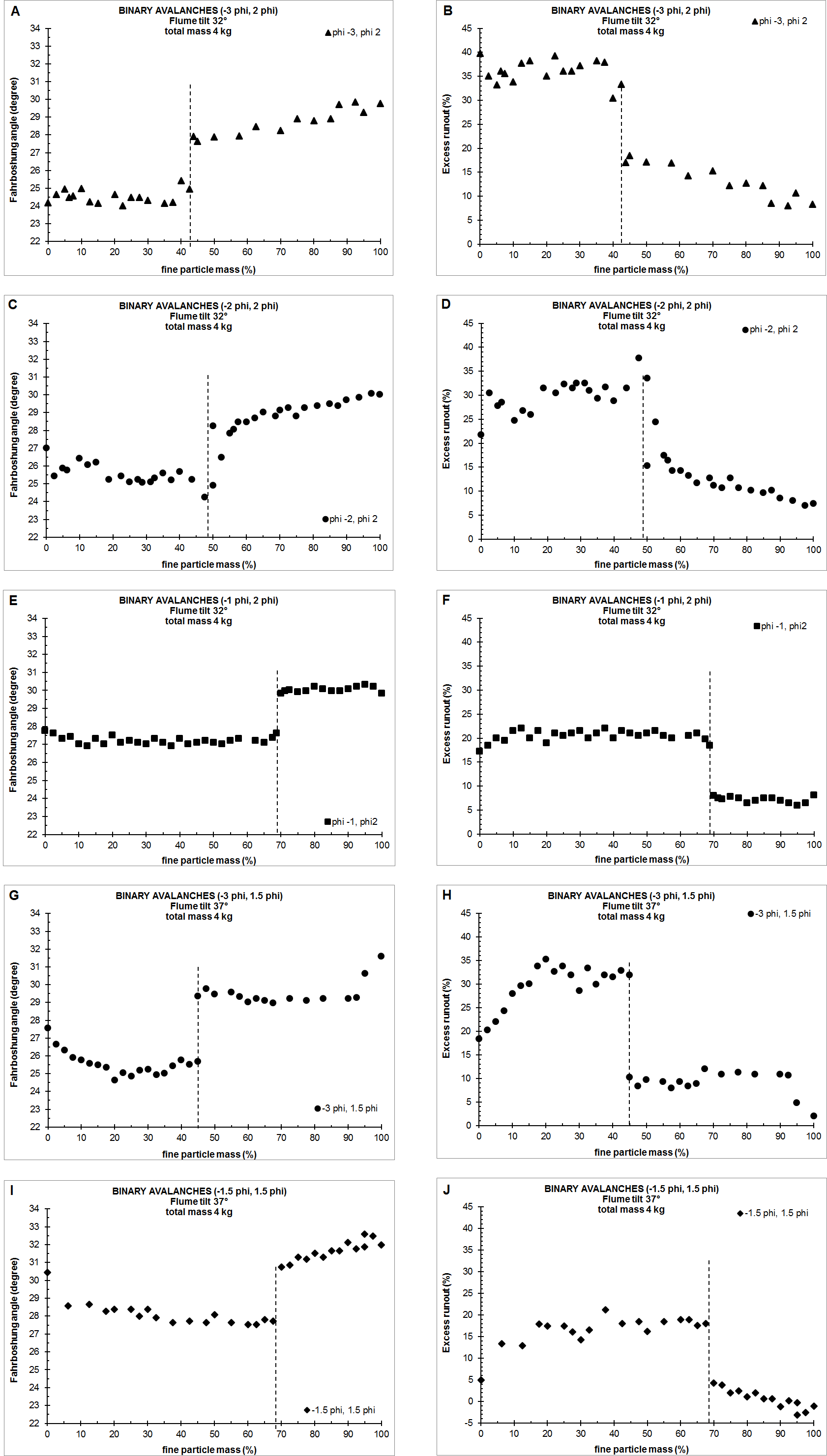}}
      \caption{\label{Fig_Fahr} (a), (c), (e), (g), and (i): Fahrböschung angle corresponding to
      avalanches containing the whole set of mixtures reported in table 1, as a function of the
      respective fine-particle content, showing the critical clustering transition pointed by the
      dashed line in each case. (b), (d), (f), (h), and (j): Excess runout (\%) for the
      above-mentioned experiment.}
    \end{center}
  \end{figure*}
  
  The excess of runout or excess of travel distance $L_e$ (Heim 1882, Hsu 1975, Corominas 1996,
  Shreve 1968, Legros 2002, Bowman \etal 2012) is the difference between the distance travelled by
  the granular flow ($L_{\max}$) and the one that may be expected ($L_f$) for a mass sliding down
  an inclined plane, in the case when a ``normal'' coefficient of friction—$\tan (32^{\circ}) =
  0.62$—is considered:
  $$
  	L_f  = \frac{H_{\max}}{\tan (32^{\circ})};
  $$
  $$
	  L_e = L_{\max} - L_f
	$$
  This friction coefficient (0.62) is expected in a purely frictional model when a block slides
  down an inclined plane, defined by Hsu (1975 and 1978). The percentage of the excess of runout
  plotted in figure~\ref{Fig_Fahr} is then:
  $$
    L_{e\%} = 100\frac{L_{\max}}{L_f} - 100.
  $$

  It can be seen that the excess runout is larger for mixtures in which the mass of large particles
  dominates over that of small particles. In order to compare the RBP of bi-disperse avalanches
  (figure~\ref{Fig_RBP}) with that of monodisperse-avalanche (figure~\ref{Fig_RBP-phi}), it is
  important to note that both sets of experiments were performed using exactly the same material
  and the same experimental flume. As it can be seen, the fine-particle content (less than a
  critical quantity) modifies the mobility (or the RBP), with a tendency of increasing it. This can
  be due to the lubrication produced by interstitial fine particles, when coarse particles are the
  largest proportion by mass, or to coarse particles dragging fine particles, when coarse particles
  are a minority compared to the fine ones by mass. Instead, with a larger-than-critical mass of
  fine grains, collisions and friction tend to slow down the avalanche trapping the coarse
  particles between the fine ones, reducing the mobility and the runout of the flow, thus acting
  like a ``sand-trap''.
  
  \section{Conclusions}
  
  We have reported the run from break point (RBP) of a binary-mixture avalanche as a function of
  its content of fine particles. Two competing behaviours, a viscous-like one, associated to fine
  particles, and an inertial one, associated to coarse particles, regulate the global behaviour of
  the mixed avalanche, depending on the proportion of each granulometric class within the mixture.
  We have found a tendency of increment of the RBP, which is produced when a small percent of fine
  particles are added to a coarse-grain avalanche, due to the lubrication among coarse grains and
  between coarse grains and the flume surface, produced by the presence of interstitial fine
  particles. This RBP increment saturates as more fine particles are added, until a sudden
  transition makes the RBP diminish abruptly to small values closer to those corresponding to the
  RBP of a monodisperse avalanche of fine particles. Here, the avalanche is dominated by a
  viscous-like behaviour, characteristic of these fine particles flowing down the inclined plane.

  The sudden transition is caused by a clustering instability and occurs when the inertial-particle
  population, which overcomes the deposit of fine or mixed particles, reaches a critical amount, at
  which a new deposit of coarse particles condenses from the dispersed cloud. By looking at those
  clusters, formed from dispersed coarse particles at the very front, we suggest that clustering
  could be at the origin of some hummock-like structures formed at the front of long-runout,
  natural rock avalanches. The variation in the content of fine particles in the avalanching
  mixture required for the transition to occur is found experimentally to be lower than 7\% by
  mass.
  
  The formation of well-defined regions of only fine and only coarse particles and a mixed,
  reverse-graded region in the deposit shows that a complete segregation can be achieved for a
  large enough inertial contrast and/or flume length. The critical content of fines at the
  transition is related to the particle-size ratio as a power law for all the studied cases.
  
  \ack
  
  We wish to thank the Geology Institute and LAIMA laboratory of Universidad Autónoma de San Luis
  Potosí, for sharing with us their facilities. This project was partially funded by CONACYT grant
  number 221961, Fondos Concurrentes UASLP and PhD-scholarship grant number 45697. We also wish to
  thank all the undergraduate students that helped us to perform the experiments.

\end{document}